\documentclass[twocolumn]{svjour3}
\pdfoutput=1
\def\makeheadbox{\relax}
\usepackage{makeidx} 
\usepackage{import}
\usepackage{graphicx}
\usepackage[font+=small,labelformat=parens,labelsep=space,skip=6pt,list=false,hypcap=false]{subcaption}
\usepackage{grffile}
\usepackage{ragged2e}
\usepackage{tabu}
\usepackage{longtable}
\usepackage[hyphens]{url}
\usepackage[usenames,dvipsnames]{xcolor}
\usepackage[breakable, theorems, skins]{./tcolorbox}
\usepackage{float}
\restylefloat{table}
\restylefloat{figure}
\usepackage{hyperref}
\usepackage[utf8]{inputenc}
\usepackage[T1]{fontenc}

\tcbset{enhanced}

\DeclareRobustCommand{\mybox}[2][gray!20]{%
\begin{tcolorbox}[   %% Adjust the following parameters at will.
        breakable,
        left=0pt,
        right=0pt,
        top=0pt,
        bottom=0pt,
        colback=#1,
        colframe=#1,
        width=\dimexpr\columnwidth\relax, 
        enlarge left by=0mm,
        boxsep=5pt,
        arc=0pt,outer arc=0pt,
        ]
        #2
\end{tcolorbox}
}

\newcommand\researchquestionformat[1]{\begin{quote}#1\end{quote}}

\newcommand\firstresearchquestion{\researchquestionformat{%
    \textbf{Research Question 1:} \emph{Have mapping studies with
      similar goals to ours been carried out?}%
  }}

\newcommand\secondresearchquestion{\researchquestionformat{%
    \textbf{Research Question 2:} \emph{What is the share of studies
      that ground their results with empirical methods?}%
  }}

\newcommand\thirdresearchquestion{\researchquestionformat{%
    \textbf{Research Question 3:} \emph{How are studies that provide
      empirical results grouped according to the ``three Vs''?  And
      what is the distribution of these different groups?}%
  }}

\newcommand\fourthresearchquestion{\researchquestionformat{%
    \textbf{Research Question 4:} \emph{What are the application areas
      of Big Data and how are they distributed?}%
  }}

\newcommand\fifthresearchquestion{\researchquestionformat{%
    \textbf{Research Question 5:} \emph{Which publication outlets are most
      prominent?}%
  }}

\newcommand\sixthresearchquestion{\researchquestionformat{%
    \textbf{Research Question 6:} \emph{Can we identify any trends
      within Empirical Big Data Research?}%
  }}

\begin{document}

%\begin{frontmatter}
%\journalname{Data Science and Engineering}
\title{Empirical Big Data Research: A Systematic Literature Mapping}

%\author{}
%\institute{
%}
\author{Bjørn Magnus Mathisen \and Leendert W. M. Wienhofen \and
  Dumitru Roman}
\institute{Bjørn Magnus Mathisen \and Leendert W. M. Wienhofen 
\at SINTEF ICT, PO Box 4760, Sluppen, NO-7465 Trondheim,
Norway\\\email{\{bjornmagnus.mathisen|leendert.wienhofen\}@sintef.no}
\and
Bjørn Magnus Mathisen
\at Department of Computer and Information Science, NTNU, NO-7491
Trondheim, Norway\\\email{bjornmm@idi.ntnu.no}
\and
Dumitru Roman
\at SINTEF ICT, P.O. Box 124 Blindern, N-0314 Oslo, Norway\\\email{dumitru.roman@sintef.no}
}

\maketitle

 \begin{abstract}
   \textit{Background:} Big Data is a relatively new field of research and
   technology, and literature reports a wide variety of concepts labeled with
   Big Data. The maturity of a research field can be measured in the number of
   publications containing empirical results. In this paper we present the
   current status of empirical research in Big Data. \textit{Method:} We
   employed a systematic mapping method with which we mapped the collected
   research according to the labels Variety, Volume and Velocity. In addition,
   we addressed the application areas of Big Data. \textit{Results:} We found
   that 151 of the assessed 1778 contributions contain a form of empirical
   result and can be mapped to one or more of the 3 V's and 59 address an
   application area. \textit{Conclusions:} The share of publications containing
   empirical results is well below the average compared to computer science
   research as a whole. In order to mature the research on Big Data, we
   recommend applying empirical methods to strengthen the confidence in the
   reported results. Based on our trend analysis we consider Volume and Variety
   to be the most promising uncharted area in Big Data.

 \end{abstract}

%\end{frontmatter}

\section{Introduction}
\label{sec:introduction}
A sharp increase in the number of publications related to the Big Data
field in the past years makes it difficult to get a good overview of
the current state-of-the-art. Big Data technology is diverse and can
be applied to many areas. Big Data features in many trend reports and
academic publications. In order to get an overview of the field, we
have performed a systematic mapping study and assessed to which degree
empirical results have been reported. In our study empirical results
mean that a technology or concept has been tested and evaluated so
that the result can be seen as a part of an evidence base. Concepts or
technology that are merely referred to and not tested or evaluated are
excluded from this study.  Generally speaking, Big Data is a
collection of large data sets with a great diversity of types so
that it becomes difficult to process by using state-of-the-art data
processing approaches or traditional data processing
platforms~\cite{PhilipChen2014}. In a 2011 Gartner
report~\cite{laney2001} Doug Laney explains the concept of Volume,
Variety and Velocity in data management. These are known as the 3V's
and characterize the concept of Big Data. In addition to these 3
fundamental V's, many other V's have emerged, though these differ per
the special feature the authors of these publications happen to need.

In 2012, Gartner revised and gave a more detailed
definition\footnote{\url{http://www.gartner.com/resId=2057415}} as:
\textit{
Big Data are high-volume, high-velocity, and/or high-variety
information assets that require new forms of processing to enable
enhanced decision making, insight discovery and process
optimization. More generally, a data set can be called Big Data if it
is formidable to perform capture, curation, analysis and visualization
on it at the current technologies.}

NIST~\cite{nistbigdata2013} defines Big Data as: 
\textit{Big data consists of advanced
techniques that harness independent resources for building scalable
data systems when the characteristics of the datasets require new
architectures for efficient storage, manipulation, and analysis.}

All agree to the fact that Big Data needs to be big, and in order to
be assessed as Big Data, one needs to address at least one of the
aspects of volume, velocity or variety. However, when one looks into
the literature, one finds quite quickly that publications that through
their title, keywords or abstract give the impression to deal with Big
Data in fact do not address these aspects. Sj{\o}berg
et al.~\cite{sjoberg2007future} state that empirical research seeks to explore,
describe, predict, and explain natural, social, or cognitive phenomena
by using evidence based on observation or experience.  It involves
obtaining and interpreting evidence by, e.g., experimentation,
systematic observation, interviews or surveys, or by the careful
examination of documents or artifacts. Work done in an empirical
manner can be used as an evidence base for further research. In order
to separate the sheep from the wool, we committed a systematic mapping
study taking into account only publications that provide empirical
results or address 3 V aspects of Big Data.

\subsection{Study approach and contribution}
\label{sec:approachhandcontribution}

During our mapping of the Big Data literature we found no systematic
review of empirical work carried out in the field of Big Data. We did
identify related studies and describe these in Section
\ref{sec:relatedsurveys}. In order to create an overview of the areas
that are addressed, this paper describes how we carried out a
systematic literature mapping with a method similar to [124] to map
existing Big Data literature with a form of empirical evidence to the
three V's of Big Data as well as application areas. The method is
described in detail in Section \ref{sec:systematicmappingprocess}. We
chose to limit the mapping to the 3 V's as these are the fundamental
issues for Big Data. Many other V-terms have been defined later though
none of these are used consistently, which limits the mapping
possibilities.

The main contributions of our study are;
\begin{itemize}
\item Systematic mapping of the findings of empirical Big Data Studies
\item Identification of Big Data studies containing empirical evidence
\item Overview of application areas of empirical Big Data Studies
\item Trend analysis of empirical Big Data Research
\item Identification of and discussion about related surveys
\end{itemize}

A summary of some of our conclusions:
\begin{itemize}
\item The number of reports on Big Data are rising, both empirical and
  non-empirical
\item Roughly 10 percent of the contributions on Big Data include empirical results
\item Application areas have been getting more attention over time
\item We recommend applying empirical methods to strengthen the
  confidence in the reported results
 \item Based on our trend analysis we consider Variety to be the most
   promising uncharted area in Big Data
\end{itemize}

\subsection{Structure of this paper}
\label{sec:structure}

The paper is structured as follows.
Section~\ref{sec:systematicmappingprocess} describes the general
method employed in the work presented in this paper as well as our
specific implementation of this method. The results of the different
stages of this work are presented in
Section~\ref{sec:results}. Section~\ref{sec:analysis} presents an
analysis of the results and Section~\ref{sec:discussion} discusses
the limitations of our study. Finally, Section~\ref{sec:conclusion}
describes our conclusion and our recommendations for further research.

%%% Local Variables:
%%% mode: latex
%%% TeX-master: "../bigdatasms"
%%% End:

\section{The systematic mapping process}
\label{sec:systematicmappingprocess}
\begin{figure}[ht!]
  \centering
    \includegraphics[width=\columnwidth]{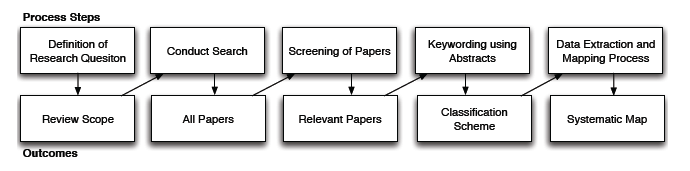}
  \caption{The stages of the systematic mapping process~\cite{Petersen2007}.}
  \label{fig:mappingprocess}
\end{figure}

The systematic mapping process is an iterative process where each step
builds upon the previous. The process starts with a research question
and ends with a systematic map, see Figure~\ref{fig:mappingprocess}.
We have based our systematic mapping procedure on Kai Petersen's and
Robert Feldt's work~\cite{Petersen2007}. In this section we first
highlight the step (in a text box) in the systematic mapping process as described by
Petersen and Feldt, each step in the process description is then
followed by a description of how we implemented this step in the
process.

\subsection{Definition of Research Questions (Research Scope)}
\label{subsec:definitionofresearchquestions}

\mybox{
The main goal of a systematic mapping study is to provide an overview
of a research area, and identify the quantity and type of research and
results available within it. Often one wants to map the frequencies of
publication over time to see trends. A secondary goal can be to
identify the forums in which research in the area has been published.
}

We have defined the following research questions:
\firstresearchquestion 
\secondresearchquestion 
\thirdresearchquestion 
\fourthresearchquestion 
\fifthresearchquestion 
\sixthresearchquestion

\subsection{Conduct Search}
\label{subsec:conductsearch}

\mybox{
The primary studies are identified by using search strings on
scientific databases or browsing manually through relevant conference
proceedings or journal publications. A good way to create the search
string is to structure them in terms of population, intervention,
comparison, and outcome \cite{Kitchenham2007}. The structure should of
course be driven by the research questions. Keywords for the search
string can be taken from each aspect of the structure. For example,
the outcome of a study (e.g., accuracy of an estimation method) could
lead to key words like ``case study'' or ``experiment'' which are research
approaches to determine this accuracy.
}

In this study we used Elsevier's
Scopus\footnote{\url{http://www.scopus.com}} for our search. Scopus
delivers the most comprehensive overview of the world's research
output in the fields of science, technology, medicine, social sciences
and arts and humanities. It claims to be the largest abstract and
citation database of peer-reviewed literature and within our domain it
is a valid choice. Test searches done within other databases all
returned subsets of the result from Scopus (e.g. the results from
``(Big and Data) and (PublishedAs:journal) and (Keywords:Big AND
Keywords:Data)'' limited to 2014 and earlier in ACM digital library
was all part of the Scopus results.)

We searched for ``Big Data'' in the title, abstract and keywords. We
included only papers that are accepted in journals, or in press for
journals as well as reviews. This resulted in the following search
string:

\begin{center}
\begin{verbatim}
TITLE-ABS-KEY("Big Data") AND DOCTYPE(ar OR re)
 AND PUBYEAR < 2015
 AND (LIMIT-TO(LANGUAGE,"English")) 
 AND (LIMIT-TO(SRCTYPE,"j")
      OR LIMIT-TO(SRCTYPE,"k"))
\end{verbatim}
\end{center}
The string above is defined by the Scopus search query language which can be
accessed at Scopus\footnote{\url{http://www.scopus.com/search/form.url?display=advanced}} where\\ ``DOCTYPE(ar OR re)'' limits to article or review,\\
``PUBYEAR < 2015'' limits to publication from before
2015, ``(LIMIT-TO(LANGUAGE,"English"))'' limits to publications with
English as the source language and
``LIMIT-TO(SRCTYPE,"j") OR LIMIT-TO(SRCTYPE,"k")'' limits to journals and
book series.

At the time of the query this also resulted in three
publications~\cite{Chen2015,Hsu2015,Wu2015} dated 2015 in direct
conflict with the query parameters. This may be because of some
journal predating an article (indexing it when it is accepted, but
before it is actually published). We have included these papers in the
data for completeness. These publications were later excluded in our selection
process and thus will not be included in any of the analysis except
from the graphs presenting the total number of publication and any
calculation or analysis that depends on the total number of
publications.

\subsection{Screening of Papers for Inclusion and Exclusion (Relevant Papers)}
\label{subsec:screeningofpapersforinclusionandexclusion}

\mybox{ Inclusion and exclusion criteria are used to exclude studies
  that are not relevant to answer the research questions. The criteria
  show that the research questions influenced the inclusion and
  exclusion criteria.  It is useful to prototype inclusion and
  exclusion parameters with a limited set of papers.}

Rather than having specific inclusion criteria other than the
selection by the search query described above, our method includes
every paper from the base corpus until excluded. Exclusion criteria
are used to exclude studies that are not relevant to answer the
research questions. One may however regard the inverse of criterion 3
and 4 as inclusion criteria. See Table~\ref{tab:criteria} for the criteria.

Given the inclusion and exclusion criteria used in the study described
in the process description, we defined criteria suitable for our
data-set. Out of the full dataset, we randomly chose 100 papers to
test the inclusion and exclusion parameters, prior to reading the full
set of papers. The random selection was based on numbering in
Endnote, we simply took approximately every 15th paper and included it
in our random set.  The first exclusion criteria is ``no abstract''
(some of the results appeared to be short-papers in magazines, and
these typically do not include abstracts), as, if there is no
abstract, we simply cannot see whether the publication is relevant or
not.  The second exclusion criteria is ``Source language other than
English''. Some abstracts were written in a way that we typically see
when a machine translation is applied, which caused doubt regarding
the actual content of the contribution. Upon further investigation of
the meta-data, we noticed that some contributions do not have English
as a source language. This will make us unable to do further
investigation when needed and therefore we chose to exclude these
articles.  Once the contribution passed the first two criteria, we
looked into the content. Only clear contributions to Big Data are of
interest for this mapping study and therefore we exclude (criterion 3)
abstracts that do not clearly define contribution of work, as well as
abstracts (criterion 4) that are clearly not related to the modern
term Big Data (e.g. \cite{Perotto1987} talks about ``Big Data
reduction'', however it is clearly not related to the modern term
``Big Data'').  Publications with very small data sets that claim that
the solution will work on a huge dataset without having a convincing
strategy for this are also excluded by criterion 4. See also
Table~\ref{tab:criteria}.  

\begin{center}
  \begin{table}[ht!] 

\centering
    
  \begin{tabu} to 1\columnwidth { | X[1,l] | X[2,l] | }
  \hline
  Exclusion criteria number & Criteria \\ \hline \hline
  1 & No abstract. \\ \hline
  2 & Source language other than English
\\ \hline
  3 & Abstract does not clearly define contribution of work. \\ \hline
  4 & Clearly not related to Big Data. \\ \hline
  \end{tabu}
  
  \caption{Exclusion criteria}
  \label{tab:criteria}
\end{table}
\end{center}

\subsection{Keywording of Abstracts (Classification Scheme)}
\label{sec:keywording}

\mybox{
Keywording is a way to reduce the time needed in developing the
classification scheme and ensuring that the scheme takes the existing
studies into account.}

We adopted the systematic process for classification from
~\cite{Petersen2007}. However, instead of searching for keywords to
base the cluster map on, in our case, the keywords were already
defined by Laney \cite{laney2001} as explained in the introduction. In
addition to the 3 V's we also defined "application area" as a keyword
to map to.  As for the mapping to empirical keywords, we extracted a
list of empirical method keywords, which has been compiled in a matrix
 showing co-occurrences in Table~\ref{tab:empiricalmethodsmatrix}.

\begin{figure}[ht!]
  \centering
    \includegraphics[width=\columnwidth]{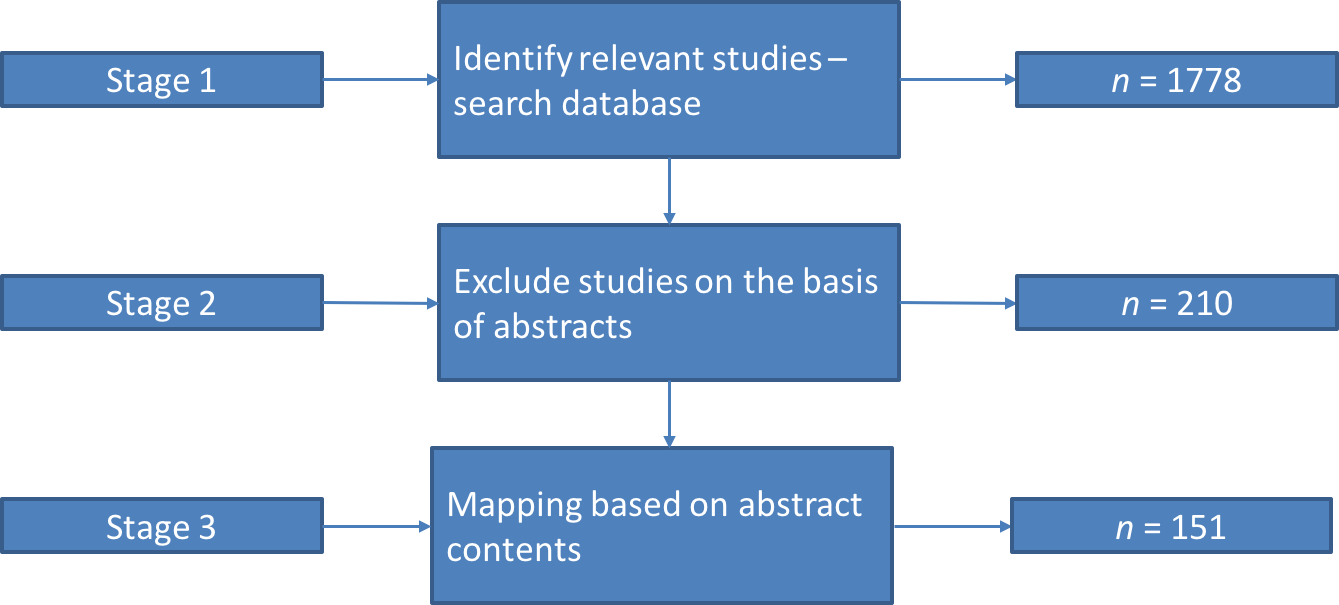}
  \caption{The three stages of the systematic mapping.}
  \label{fig:mapping_stages}
\end{figure}

\subsection{Data Extraction and Mapping of Studies (Systematic Map)}
\label{sec:mapping}

\mybox{
When having the classification scheme in place, the relevant articles
are sorted into the scheme, i.e., the actual data extraction takes
place. The classification scheme evolves while doing the data
extraction, like adding new categories or merging and splitting
existing categories. A scheme, for example in Excel, should be used to
document the data extraction process. The table should contain each
category of the classification scheme. When the reviewers enter the
data of a paper into the scheme, they provide a short rationale why
the paper should be in a certain category (for example, why the paper
applied evaluation research). From the final table, the frequencies of
publications in each category can be calculated.

Mapping data in graphs is a useful aid for the reader to understand
the analysis. Visualization alternatives could be found in statistics,
HCI and information visualization fields.  }

We began by categorizing all articles based on their abstracts into
four categories. We determine whether the article in question have
contributed to the Big Data field itself in term of either volume,
variety or velocity. If the article is simply applying one or more Big
Data techniques in a case, we identified whether this is a Big Data
experiment that has contributed to the field itself by proving that
``doing X is possible with these techniques'' to a reasonable degree
or if this has little effect on the field. In addition, we categorized the
contribution according to the empirical methods used:

\begin{enumerate}
\item Volume: Describes improvements and progress within technologies
  and methods for handling increases to the volume of
  data. E.g. Optimizing analysis methods to reduce runtime ($O(N^3)
  \rightarrow O(N^2)$) thus enabling users to handle larger volume of
  data, but not approaching stream/real-time speed (velocity), or
  improved methods for handling storage and transfer of Big Data.
\item Variety: Describes improvements and progress within technologies
  for handling variety of data. E.g. new methods for classification
  that exploits very large amounts of data.
\item Velocity: Describes improvements and progress within technologies for
  coping with the speed of incoming data. E.g. Decreasing turnaround/response
  time for analysis, approaching real-time/stream analysis or completion time
  guarantees. Technologies and methods for handling a fire hose of incoming
  data\footnote{See
    \url{http://cacm.acm.org/blogs/blog-cacm/155468-what-does-big-data-mean/fulltext}},
  one pass algorithms for computing approximate
  statistic/analytics\cite{leskovec2014mining}
%  the method/analysis more applicable for stream analysis/real-time analysis.
\item Application area: Describes Big Data technology 
  different application areas; not innovating through new Big Data
  technology but through applying Big Data Technology to new areas.
\end{enumerate}

%%% Local Variables:
%%% mode: latex
%%% TeX-master: "../bigdatasms"
%%% End:

\section{Findings from Data}
\label{sec:results}

This section describes the outcomes of the steps in the method described
in Section~\ref{sec:systematicmappingprocess}; The results chapter map
directly to the stages of the systematic mapping process described in
Figure~\ref{fig:mappingprocess}.

% The following section is structured as follows.
% Subsection~\ref{sec:reviewscope} describes the result of ``Definition
% of research question''. Subsection~\ref{sec:allpapers} describes the
% result of ``Conduct Search''. Subsection~\ref{sec:relevantpapers}
% describes the result of ``Screening of papers''.
% Subsection~\ref{sec:classificationscheme} describes the result of
% ``Keywording using abstracts''. Subsection~\ref{sec:systematicmap}
% describes the result of ``Data Extraction and Mapping Process''

The following section is structured as follows.
Subsection~\ref{sec:reviewscope} describes the result of Definition
of research question, conducting the search and screening of papers.
Subsection~\ref{sec:classificationscheme} describes the result of
``Keywording using abstracts''. Subsection~\ref{sec:systematicmap}
describes the result of ``Data Extraction and Mapping Process''

The method used can be summarized in three main stages, as shown in
Figure~\ref{fig:mapping_stages}. 

The outcome of stage 3 is described in Section~\ref{sec:systematicmap}.

\subsection{Review scope, all papers and relevant papers}
\label{sec:reviewscope}
\textbf{Review scope:}
The goal of this study is to understand the state-of-the-art within
the field of Big Data. We aim to identify past and current trends.  A
secondary goal is to identify the forums that publishes research in
the field of study. Our research questions reflects these goals.

\begin{figure}[ht!]
  \centering
    \includegraphics[width=\columnwidth]{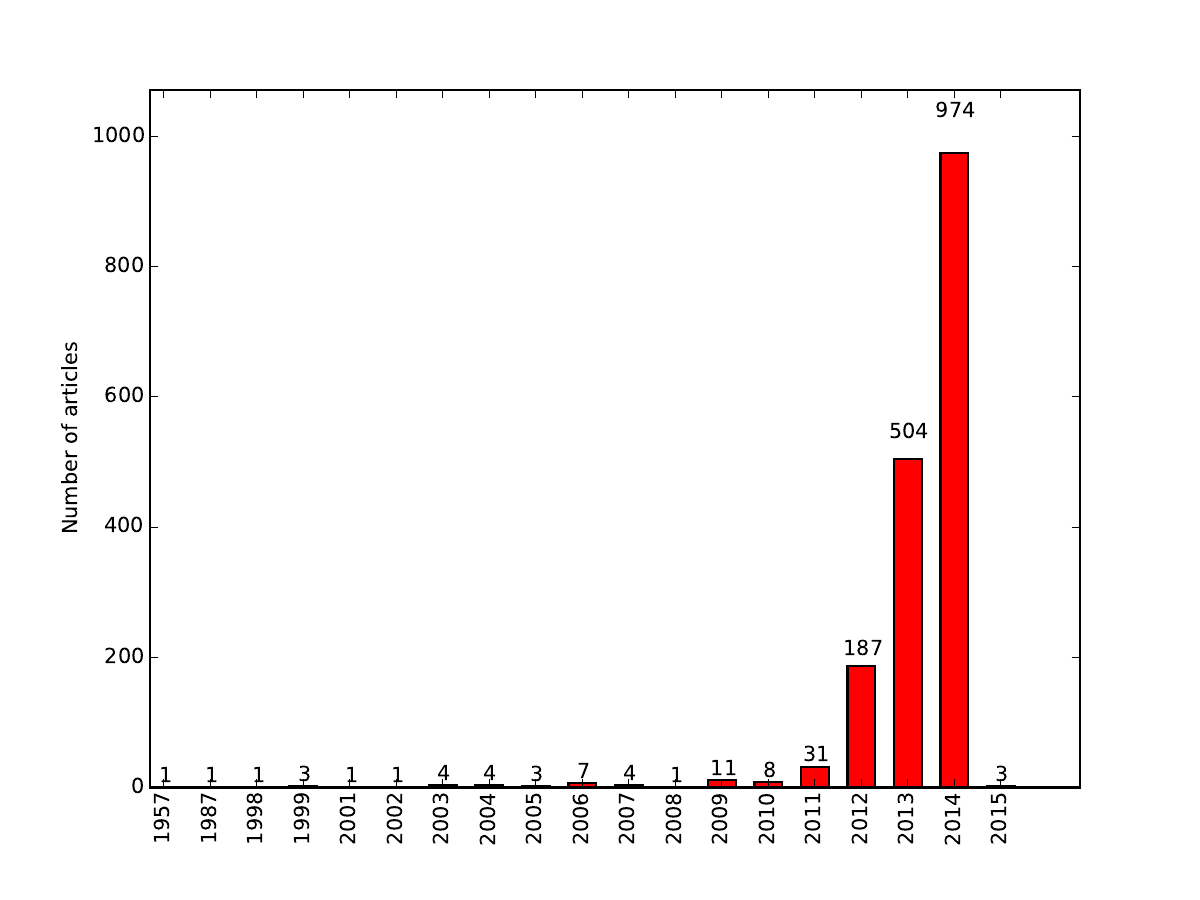}
    \caption{Distribution of the journal articles that contains ``Big
      Data'' or ``Big Datum'' in title, abstracts or keywords according to
      published year. This includes old uses of the term ``Big Data''
      as can be seen by the publication dated 1957.}
  \label{fig:allyearhistogram}
\end{figure}

\textbf{All papers:} The query as described in
Subsection~\ref{subsec:conductsearch} was sent to Elsevier's SCOPUS February
12th 2015, and resulted in 1778 publications. However some of the publications
had duplicate entries in the Scopus database, typically registered in two
different years or in two different publications. These duplicate entries were
removed and the final number of unique publications is 1749. In
Figure~\ref{fig:allyearhistogram} we have listed the distribution of
publications per year.

\begin{figure}[ht!]
  \centering
    \includegraphics[width=\columnwidth]{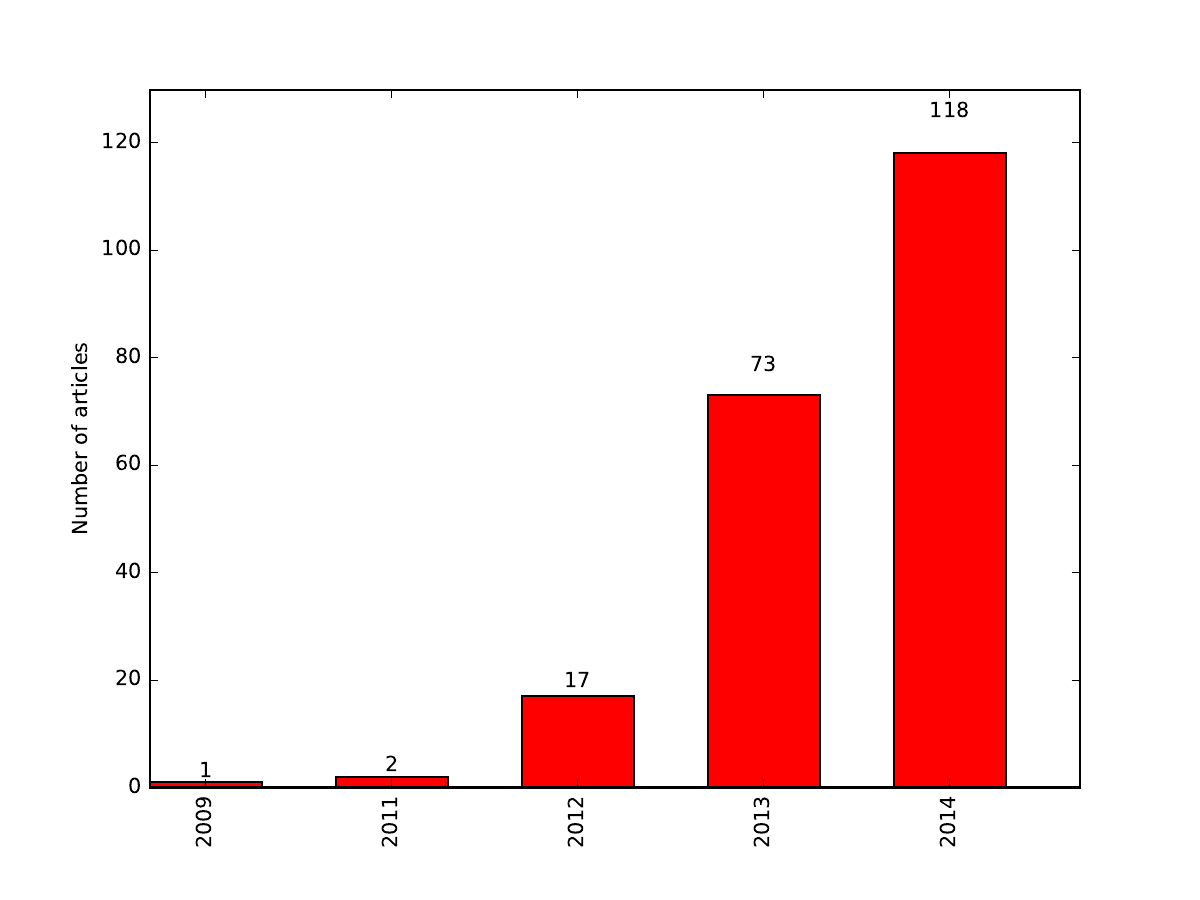}
    \caption{Distribution of the included journal articles according to
      published year.}
  \label{fig:5staryearhistogram}
\end{figure}

\textbf{Relevant papers:}
In Figure~\ref{fig:5staryearhistogram} we show the number of
included papers per year. We also present which of
the journals were most prominent in Table~\ref{tab:journals}.

\begin{table}[H]
  \begin{center}
    \begin{scriptsize}
      \begin{tabu} to \columnwidth { X[l,3] | X[l,1.5] }
        \hline
        Publication source & No.\\
        \hline
        Lecture Notes in Computer Science & 13\\
        Proceedings of the VLDB Endowment & 8\\
        Future Generation Computer Systems & 7\\
        IEEE Transactions on Emerging Topics in Computing & 6\\
        Distributed and Parallel Databases & 5\\
        IEEE Network & 4\\
        Expert Systems with Applications & 4\\
        IEEE Transactions on Knowledge and Data Engineering & 4\\
        Journal of Supercomputing & 4\\
        Knowledge and Information Systems & 4\\
        %% International Journal of Distributed Sensor Networks & 3\\
        %% Big Data Research & 3\\
        %% PLoS ONE & 3\\
        %% International Journal of Multimedia and Ubiquitous Engineering & 3\\
        %% International Journal of Communication Systems & 3\\
        %% Parallel Computing & 3\\
        %% Journal of Internet Technology & 3\\
        %% Cluster Computing & 3\\
        %% IEEE Transactions on Parallel and Distributed Systems & 3\\
        %% Neural Networks & 2\\
        %% International Journal of Approximate Reasoning & 2\\
        %% NTT Technical Review & 2\\
        %% Studies in Computational Intelligence & 2\\
        %% Computing & 2\\
        %% Journal of Parallel and Distributed Computing & 2\\
        %% ACM Transactions on Knowledge Discovery from Data & 2\\
        %% Pattern Recognition Letters & 2\\
        %% Performance Evaluation & 2\\
        %% Computer Science and Information Systems & 2\\
        %% Tsinghua Science and Technology & 2\\
        %% International Review on Computers and Software & 2\\
        %% International Journal of Business Process Integration and Management & 2\\
        %% Concurrency Computation Practice and Experience & 2\\
        %% Machine Learning & 2\\
        \hline
        \end{tabu}
    \end{scriptsize}
  \end{center}
  \caption{Publication sources (labeled as journals by Scopus) by number of included publications.}
  \label{tab:journals}
\end{table}

\subsection{Classification scheme}
\label{sec:classificationscheme}
\begin{figure}[ht]
  \centering
    \includegraphics[width=\columnwidth]{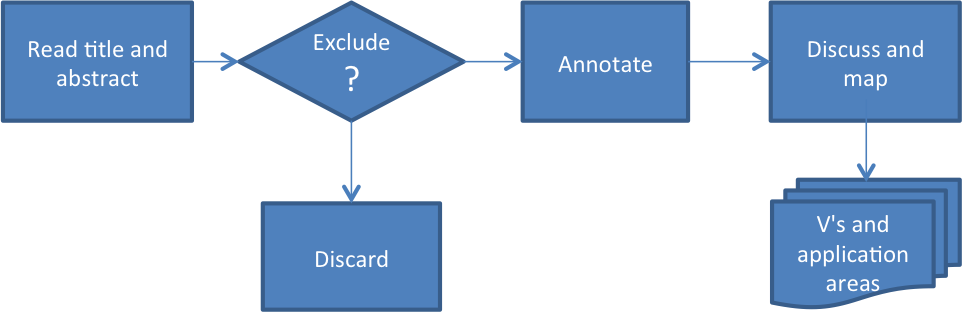}
    \caption{A flowchart describing the classification scheme applied
      in this review process.}
  \label{fig:classificationscheme}
\end{figure}

First we produced our primary corpus by applying the exclusion
criteria (see Table~\ref{tab:criteria}) to the initial population of
papers described in the previous section. When reading the title and
abstract we first checked if the paper was affected by any of our
exclusion parameters. After this check was passed, keywording was
applied in the sense that main contributions were highlighted in the
meta-data. This process is outlined in
Figure~\ref{fig:classificationscheme}. 
Some articles turned out to
be application area descriptions rather than contributions to any of
the V's. These were classified accordingly.

\subsection{Systematic map}
\label{sec:systematicmap}
\begin{figure}[H]
  \centering
    \includegraphics[width=\columnwidth]{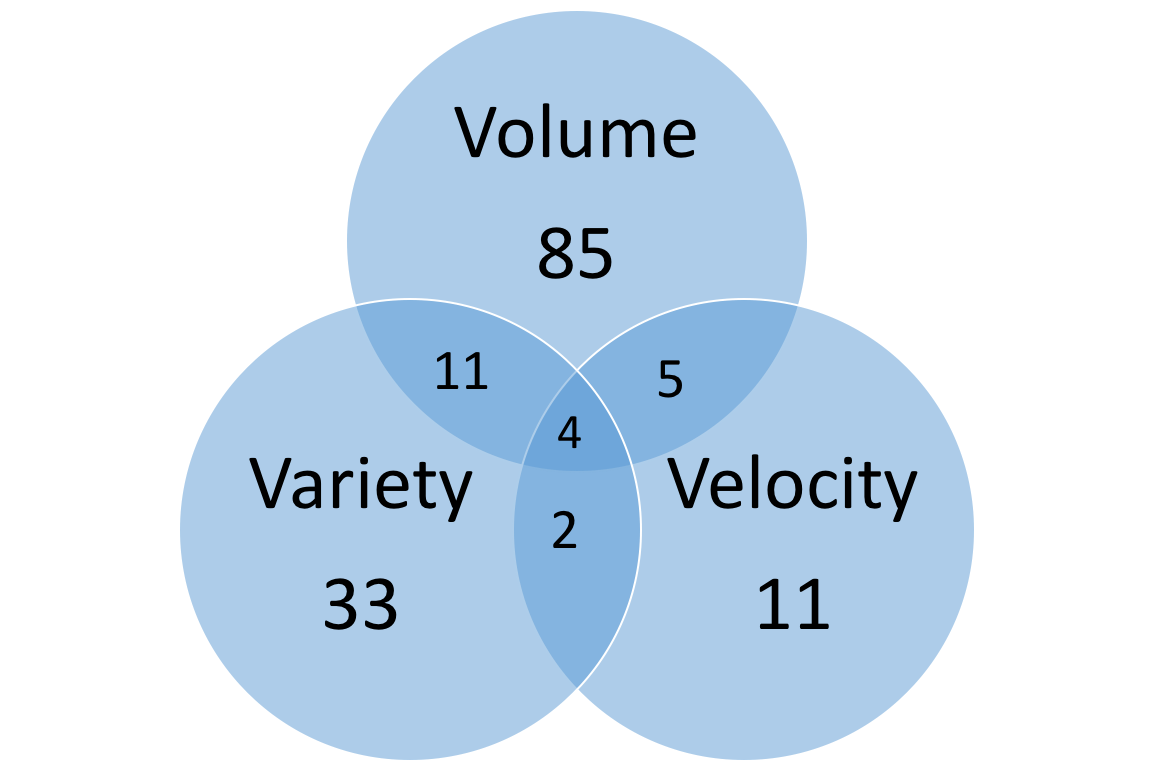}
  \caption{Venn diagram of the mapped studies.}
  \label{fig:venn_diagram}
\end{figure}

Through our classification we mapped the publications onto four
categories Volume, Velocity, Variety and Application area according to
what they contributed with in the field of Big
Data. Table~\ref{tab:threevyear} shows the mapping results according
to publication year. The publications are also mapped onto a Venn
diagram in Figure~\ref{fig:venn_diagram}, showing the number of
publications in each V and their intersections. The numbers in
Figure~\ref{fig:venn_diagram} correspond to Table~\ref{tab:vtable}
which also includes the references to the publications. Application
areas are listed in Table~\ref{tab:applicationarea}.

\subsection{Empirical Methods}
\label{sec:empiricalmethods}
In Table~\ref{tab:empiricalmethodsmatrix} we provide an overview of
the means that were identified as methods for being evaluated as being
empirical. In the cross matrix we see that some contribution use
several methods in order to prove the value of their work.

\begin{table*}[t]\centering  
  \begin{center}
    \begin{tabu} to \linewidth { X[l,3] | X[l,1.5] | X[l,1] | X[l,1.5] |
        X[l,1] | X[l,1] | X[l,0.7] | X[l,1] | X[l,1] | X[l,1] | X[l,1]}
      \hline
      & Bench. & Case study & Demon. & Eval. & Exp. & Vali. & Impl. & Model & Simul. & Verify\\
      \hline
      Benchmark & 14 & 0 & 4 & 0 & 8 & 1 & 4 & 0 & 0 & 1\\
      \hline
      Case study & 0 & 6 & 1 & 1 & 0 & 0 & 0 & 0 & 0 & 0\\
      \hline
      Demonstrate & 4 & 1 & 48 & 10 & 32 & 3 & 2 & 3 & 2 & 1\\
      \hline
      Evaluate & 0 & 1 & 10 & 35 & 14 & 1 & 4 & 3 & 1 & 1\\
      \hline
      Experiment & 8 & 0 & 32 & 14 & 108 & 7 & 13 & 4 & 3 & 4\\
      \hline
      Implement & 4 & 0 & 2 & 4 & 13 & 1 & 22 & 1 & 1 & 1\\
      \hline
      Model & 0 & 0 & 3 & 3 & 4 & 4 & 1 & 23 & 1 & 0\\
      \hline
      Simulation & 0 & 0 & 2 & 1 & 3 & 1 & 1 & 1 & 11 & 0\\
      \hline
      Validate & 1 & 0 & 3 & 1 & 7 & 12 & 1 & 4 & 1 & 0\\
      \hline
      Verify & 1 & 0 & 1 & 1 & 4 & 0 & 1 & 0 & 0 & 8\\
      \hline
      \hline
      Total & 32 & 8 & 106 & 70 & 193 & 30 & 49 & 39 & 20 & 16\\
      \hline
    \end{tabu}
  \end{center}
  \caption{Cross matrix of empirical methods, showing which empirical
    method keyword was used how many times, and their co-occurrence. The
    top row is an abbreviated version of the first column.}
  \label{tab:empiricalmethodsmatrix}
\end{table*}

\begin{table}[H]
  \centering
  \begin{center}
    \begin{tabular}{r|rrr}
      \hline
      Reference Year & Variety & Velocity & Volume\\
      \hline
      2009 & 1 & 0 & 1\\
      2011 & 0 & 0 & 2\\
      2012 & 1 & 3 & 9\\
      2013 & 19 & 8 & 39\\
      2014 & 28 & 12 & 54\\
      \hline
      Total & 49 & 23 & 105\\
      \hline
    \end{tabular}
  \end{center}
  \caption{Three V's according to publishing year.}
  \label{tab:threevyear}
\end{table}

\begin{table}[H]\centering
  \begin{center}
    \begin{tabular}{l|rrrr}
      \hline
      & 2014 & 2013 & 2012 & 2011\\
      \hline
      Benchmark & 8 & 5 & 1 & 0\\
      Case study & 4 & 2 & 0 & 0\\
      Demonstrate & 25 & 20 & 2 & 1\\
      Evaluate & 20 & 10 & 4 & 1\\
      Experiment & 67 & 37 & 4 & 0\\
      Implement & 14 & 5 & 2 & 1\\
      Model & 10 & 12 & 1 & 0\\
      Simulation & 5 & 4 & 2 & 0\\
      Validate & 7 & 4 & 1 & 0\\
      Verify & 4 & 2 & 1 & 1\\
      \hline
      Total & 164 & 101 & 18 & 4\\
      \hline
    \end{tabular}
  \end{center}
  \caption{Empirical methods according to reference year.}
  \label{tab:empiricalmethodsyear}
\end{table}

%%% Local Variables:
%%% mode: latex
%%% TeX-master: "../bigdatasms"
%%% End:

\section{Analysis}
\label{sec:analysis}

From the total of 1749 included studies, Figure
\ref{fig:5staryearhistogram} depicts the distribution of the included
studies sorted by publication year. Starting with a total of 3 papers
in 2009-2011, we see that in 2012 there was an increase in relevant
publications and near fourfold in 2013, this trend continues into
2014. So, we can state there is a clear up-going trend in relevant
publications.

Of the 210 included studies, 151 could be mapped onto one or more of
the V's, the remaining 59 are papers describing Big Data technologies
applied to application areas. In the VENN diagram we chose to exclude
to view the application areas and keep the focus on the V's. The most
addressed area is volume with 85 publications, followed by variety
with 33 and velocity with 11. Research addressing both variety and volume is
most prominent with 11 included contributions, whereas volume and velocity
is combined 5 times. Finally, variety and velocity
have five included contributions and only four studies
%\cite{Ding2014,Ferrera2013,Liang2014a,Zhao2013}
\cite{Ding2014,Ferrera2013,Lasalle2014,Liang2014a} mention all three of the main
areas of Big Data. From this we can conclude that the most mature areas in terms
of published results are Velocity and Volume. We do want to note that many of
the contributions mention Hadoop and MapReduce as a basis platform while the
focus of content is directed towards velocity and/or variety. This may indicate
that the storage is taken for granted when this is used.

\begin{table*}[t]
  \begin{center}
    \begin{tabu} to \linewidth { X[l,1] X[c,1] X[l,3]}
      \hline
      V & Publication count & Publications\\
      \hline

      Variety & 33 &
      \cite{Abawajy2013}\cite{Bantouna2013}\cite{Bendler2014}\cite{Cai2014}\cite{Chen2013b}
      \cite{Chen2014b}\cite{Farrahi2013}\cite{Gong2014}\cite{Guang-Ming2013}\cite{He2014}\cite{Kim2013}
      \cite{Kuang2013}\cite{Kwon2013}\cite{Li2014}\cite{Li2014b}\cite{Li2014c}
      \cite{Liu2014a}\cite{Liu2014e}\cite{Lomotey2014}\cite{Lu2013}\cite{Ma2014b}
      \cite{Muja2014}\cite{Narducci2013}\cite{Papageorgiou2014}\cite{Park2013}\cite{Peng2013}
      \cite{Silva2014}\cite{Steffene2014}\cite{Tan2014}\cite{Villa2014}\cite{Wang2014b}\cite{Yong2014}
      \cite{Zhu2014}

      \\
      \hline

      Volume & 85 &
      \cite{Ahmadi2013}\cite{Ai-Mei2014}\cite{Alghamdi2014}\cite{Brandau2013}\cite{Campa2014}
      \cite{Cano2014}\cite{Casu2014}\cite{Chang2014}\cite{Chen2012}\cite{Chen2013}\cite{Chen2014a}
      \cite{Chen2014}\cite{Costan2013}\cite{Cui2014}\cite{Ding2013}\cite{Djenouri2014}\cite{Ewen2012}
      \cite{Fan2013}\cite{Floratou2012}\cite{Gu2013a}\cite{Gu2014b}\cite{Gu2014}\cite{Gugnani2014}
      \cite{Guo2014a}\cite{Han2013}\cite{Hasan2014}\cite{He2014a}\cite{Herodotou2011}
      \cite{Hueske2012}\cite{Ibrahim2013}\cite{Iwashita2014}\cite{Jardak2014}\cite{Ji2014a}
      \cite{Kc2014} \cite{Khakhutskyy2014} \cite{Kim2014a}
      \cite{Laptev2012}\cite{Lee2013} \cite{Li2013} \cite{Liang2014}
      \cite{Lin2013a} \cite{Liu2013} \cite{Liu2013c}
      \cite{Luo2013}\cite{Ma2014a} \cite{Maeda2014}
      \cite{Majkowska2013} \cite{Mohamed2013} \cite{Niu2013}
      \cite{Ordonez2014} \cite{Park2014} \cite{Richter2014}
      \cite{Slagter2013}\cite{Su2014} \cite{Suchard2013}
      \cite{Ting2013} \cite{Tretyakov2013} \cite{Venkitaramanan2014}
      \cite{Wang2013} \cite{Wang2013a} \cite{Wang2014a}
      \cite{Wang2014} \cite{Washisaka2011} \cite{Xia2013}
      \cite{Xia2014b}\cite{Xia2014a}\cite{Xia2014} \cite{Xie2013}
      \cite{Xiao2012} \cite{Xiao2014} \cite{Xu2014} \cite{Yan2013}
      \cite{Yang2013b} \cite{Yang2014} \cite{Yao2014}
      \cite{Yi2014}\cite{Yin2014} \cite{Zhang2012} \cite{Zhang2013d}
      \cite{Zhang2013c}\cite{Zhang2014b} \cite{Zhang2014b2}
      \cite{Zhao2014} \cite{Zhong2013} \cite{Zong2014}
      \\
      \hline

      Velocity & 11 &

      \cite{Ge2014}\cite{Gu2014a}\cite{Hu2014}
      \cite{Hunter2013}\cite{Jayalath2014}\cite{Lam2012}
      \cite{Seo2014a}\cite{Verma2012}\cite{Yang2013}\cite{Ye2013}
      \cite{Zhang2014d}
      \\
      \hline

      Volume and Velocity & 5 &
      \cite{Abad2013}\cite{Aniello2014}\cite{Lin2013}\cite{Oda2012}
      \cite{Zou2014}
      \\
      \hline

      Velocity And Variety & 2 &
      \cite{Guo2014}\cite{Yang2013a}
      \\
      \hline

      Variety and Volume & 11 &
      \cite{Alexandrov2014}\cite{Feng2013}\cite{GuerAli2013}\cite{Havens2012}\cite{Liu2013a}\cite{Sapin2014}
      \cite{Song2013}\cite{Stencl2009}\cite{Sun2013}\cite{Tang2014}\cite{Zhao2013}
      \\
      \hline

      Volume, Variety \& Velocity & 4 &
      \cite{Ding2014}\cite{Ferrera2013}\cite{Lasalle2014}\cite{Liang2014a}
      \\
      \hline
    \end{tabu}
  \end{center}
  \caption{Number of publications classified to that V
    and the references to those publication.}
  \label{tab:vtable}

\end{table*}

%\section{Studies not classified as a V}
%\label{sec:studiesnotv}

% From 31st of October 2013 we only had:

% Adaptive learning \cite{Li2012},
% Basket Analysis \cite{Videla-Cavieres2014}, Benchmarking per joule and
% second \cite{Luo2012}, Campus location analysis \cite{Caballe2013},
% Decreasing complexity \cite{Ballings2012}, Image and video
% classification tasks \cite{Faria2013}, Information Extraction
% \cite{Powell2012}, Intelligent Transport Systems \cite{Fiosina2013},
% Optical Character Regocnition \cite{Hong2013}, Privacy preservation \cite{Zhang2013a},
% Results based on Big Data technology \cite{Millie2013}, Risk Analysis
% \cite{Miller2013}, Trend Discovery \cite{Hrovat2013},
% Twitter emotions \cite{Bliss2012}, Twitter geographical analysis
% \cite{Crampton2013}, Visualisation \cite{Yang2013,Glatz2013}

\begin{table}[H]
  \begin{center}
    \begin{tabular}{r|r|r|l}
      \hline
      & total & included & \%\\
      \hline
      before 2009 & 31 & 0 & 0,00\\
      2009 & 11 & 1 & 9,09\\
      2010 & 8 & 0 & 0,00\\
      2011 & 31 & 2 & 6,45\\
      2012 & 187 & 17 & 9,09\\
      2013 & 504 & 73 & 14,48\\
      2014 & 974 & 118 & 12,11\\
      2015 & 3 & 0 & 0,00\\
      \hline
      sum & 1749 & 211 & 12,01 \\
      \hline
    \end{tabular}
  \end{center}
  \caption{Total number of journal publications per year.}
  \label{tab:totalpublications}
\end{table}

\begin{figure}[H]
  \centering
    \includegraphics[width=\columnwidth]{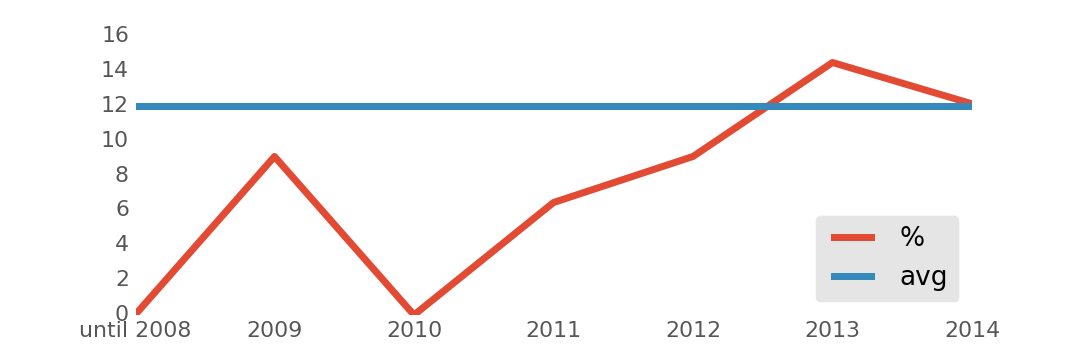}
    \caption{Percentage of empirical (included) studies per year.}
  \label{fig:includedstudiespercentagegraph}
\end{figure}

\begin{figure}[H]
  \centering
    \includegraphics[width=\columnwidth]{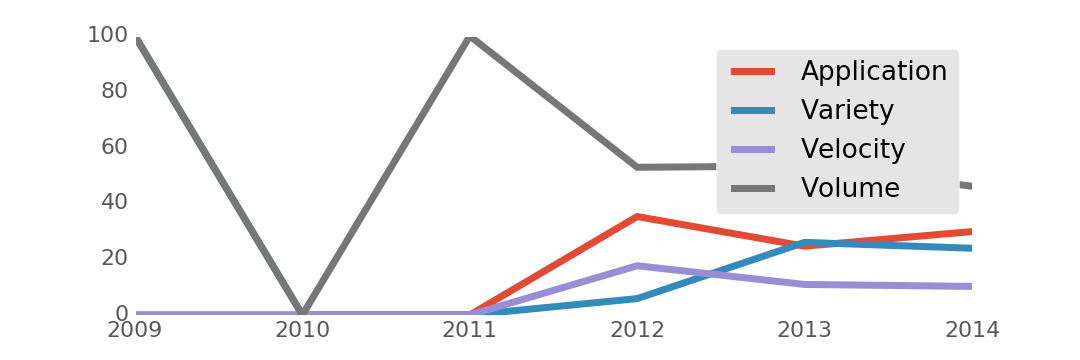}
    \caption{Percentage of empirical (included) studies per year per
      mapped region: Volume, Velocity, Variety and Application Area.}
  \label{fig:inclusionpercentagepermappedregion}
\end{figure}

\begin{figure}[H]
  \centering
    \includegraphics[width=\columnwidth]{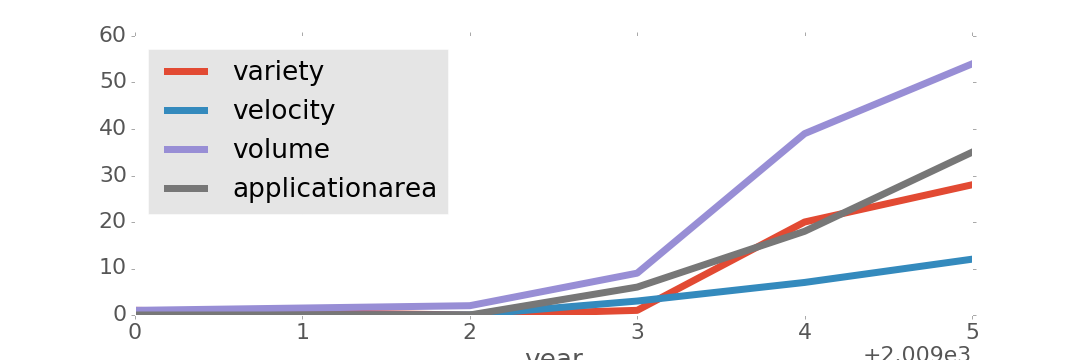}
    \caption{Number of empirical (included) studies per year per
      mapped region: Volume, Velocity, Variety and Application Area.}
  \label{fig:includedstudiesgraph}
\end{figure}

Table~\ref{tab:totalpublications} and
Figure~\ref{fig:inclusionpercentagepermappedregion} give an overview of the
total number of journal papers per year that we have assessed, as well as the
number of included paper. It becomes very clear that the majority of
publications do not have empirical findings. These numbers are also presented in
Figure~\ref{fig:5staryearhistogram} (included papers per year) and
Figure~\ref{fig:allyearhistogram} (total papers per year). The inclusion
percentage can be seen graphed in
Figure~\ref{fig:includedstudiespercentagegraph} and also per V and Application
area in Figure~\ref{fig:includedstudiesgraph}

We have mapped the included studies to 4 categories, Variety,
Velocity, Volume and Application area. The latter means that a paper
does contribute empirically to Big Data by applying Big Data
technology to a field, however, without forwarding the technology
itself.  Some papers address multiple V's, and if so, are mapped
accordingly. Hence, the total number of V's is higher than the total
number of included papers.

\begin{table*}[t]
  \begin{center}
    \begin{tabu} to \linewidth { X[r,0.4]|X[r,0.65]|X[l,0.95]|X[r,0.65]|X[l,0.95]|X[r,0.65]|X[l,0.95]|X[r,1]|X[l,0.95]}
      \hline
      & Variety & \%    change & Velocity & \%  change & Volume & \%
      change & Application & \%  change\\
      \hline
      2009 & 1 & N/A & 0 & N/A & 1 & N/A & 0 & N/A\\
      2010 & - & N/A & - & N/A & - & N/A & - & N/A\\
      2011 & 0 & N/A & 0 & N/A & 2 & N/A & 0 & N/A\\
      2012 & 1 & N/A & 3 & N/A & 9 & 450 & 6 & N/A\\
      2013 & 19 & 1900 & 8 & 266,66 & 39 & 433,33 & 18 & 300\\
      2014 & 28 & 147,36 & 12 & 150 & 54 & 138,46 & 35 & 194,44\\
      \hline
    \end{tabu}
  \end{center}
  \caption{Included publications and their mapping to V and application
    area. The table also reflects the change over time in percentage.}
  \label{tab:mappingvsandapplicationareas}
\end{table*}

Table~\ref{tab:mappingvsandapplicationareas} gives an overview of the included
papers mapped to according to the V and application area. The column ``\%
change'' indicates the change from the year before and is meant to give an
indication on whether there is growth. We see that in the past 3 years there has
been an increase for all categories in absolute numbers. Though, measured in
percentage growth compared to the previous year we see a decline.

\begin{table*}[t]
  \begin{center}
    \begin{tabu} to \linewidth {X[r] | X[r] | X[r] | X[r] | X[r] |
        X[r] | X[r] | X[r] | X[r]}
      \hline
      & Variety & included \% & Velocity & included \% & Volume & included \% & Appl. & included  \%\\
      \hline
      2009 & 1 & 100,0 & 0 & 0,00 & 1 & 100,0 & 0 & 0,00\\
      2010 & - & N/A & - & N/A & - & N/A & - & N/A\\
      2011 & 0 & 0,0 & 0 & 0,0 & 2 & 100,0 & 0 & 0,0\\
      2012 & 1 & 5,88 & 3 & 17,65 & 9 & 52,94 & 6 & 35,29\\
      2013 & 19 & 26,03 & 8 & 10,95 & 39 & 53,42 & 18 & 24,66\\
      2014 & 28 & 23,93 & 12 & 10,26 & 54 & 46,15 & 35 & 29,91\\
      \hline
    \end{tabu}
  \end{center}
  \caption{Overview of the included publications mapped to V's and the
    applications areas.}
  \label{tab:inclusionmappingvsandapplicationareas}
\end{table*}

Table~\ref{tab:inclusionmappingvsandapplicationareas} gives an overview of the
included papers mapped according to the V's and application areas. The column
``included \%'' indicates the percentage of included papers. For example, in
2009 we included 1 paper which was mapped on both Variety and Volume (also
explaining why the total can be higher than 100\% per year). In the past 3
years, for Variety we see that the percentage of included papers has increased
quite a bit from 2012 to 2013, with a little decline to 2014. The inclusion
percentage of Volume is stable from 2012 to 2015. Velocity seems to drop from
2012 to 2013 and stabilize to 2014, but the amount of publications in 2012 is so
low that it is not statistically significant. Though, the total number of
publications is still increasing.

59 of the included studies is not classified as a direct contributor
to any of the three V's, rather the publication describes Big Data
technology applied to different domains to such a degree that it can
be viewed as a contribution to Big Data as a field.

\begin{table}[h!]
  \centering
  \begin{tabu} to 0.9\columnwidth {  X[2.5,l] X[0.4,l] X[1.5,l] }
    Application area & \#  & Publications \\ \hline
    Social (network) analysis & 8 &
    \cite{Bliss2012,Crampton2013,Gross2014,Guan2014,Jiang2014,Knudsen2014,Lu2014,Corley2012} \\

    (Cyber) Security and privacy & 6 &
    \cite{Kim2014,Liu2014,Xu2014a,Zhang2013a,Zhang2014a,Zhou2014} \\ 

    Visual analytics & 5 &
    \cite{Glatz2013,Hrovat2013,Liu2014b,Ma2014,Yang2013} \\

    Predictive analytics & 4 &
    \cite{Ballings2012,Chang2014a,Zhang2014,Zhong2014} \\

    Intelligent Transport Systems & 4 &
    \cite{Dobre2014,Fiosina2013,Liu2014c,Zhang2013e} \\

    Search engine/data exploration & 3 & 
    \cite{Kitsos2014,Lee2014,Qin2014} \\

    Environmental monitoring and management & 3 &
    \cite{Fang2014,Hassan2014,Tang2014a} \\

    (Bio)Medical & 3 & 
    \cite{Jayapandian2013,Liu2014d,Mancini2014} \\

    Text Extraction & 3 &
    \cite{Hong2013,Ji2014,Powell2012} \\
    \hline
  \end{tabu}
  \caption{Publications grouped by application area.}
  \label{tab:applicationarea}
\end{table}

In addition we have studies within
Recommendations~\cite{Meng2014,Seo2014}, Cost
reduction~\cite{Chien2014}, Image and video classification
tasks~\cite{Faria2013}, Stimulation of learning
experience~\cite{Caballe2013}, Clustering~\cite{Fahad2013},
ATC~\cite{Hurter2014}, Telecom~\cite{Jun2013}, Cloud~\cite{Luo2012},
Kernel spectral clustering~\cite{Mall2013,Mall2014} (However these were
very close to be classified as applicable to Velocity and
Variability), Knowledge provision~\cite{Millie2013}, Smart
Grid~\cite{Simmhan2013}, Analytics~\cite{Sun2014}, Space~\cite{Tapiador2014},
Criminal investigation~\cite{Banerveld2014},
Marketing~\cite{Videla-Cavieres2014}, retrieval of learning
objects~\cite{Vimal2013}, Bibliometrics~\cite{Xian2014}, Service
operation~\cite{Li2012}, recreational studies~\cite{Wood2013}

We cannot give a conclusive trend analysis based on our study, though
as we do have indications, we wanted to see if these coincide with
generally available trend reports.  Trend reports and predictions are
abundant; a quick Google search on ``latest trends in Big Data'' returns
millions of results. At the time of search (on Google), the first hits
were:

Gartner Predicts Three Big Data Trends for Business
Intelligence\footnote{\url{http://www.forbes.com/sites/gartnergroup/2015/02/12/gartner-predicts-three-big-data-trends-for-business-intelligence/}}:

\begin{table}[H]
  \begin{center}
    \begin{tabu} to \columnwidth { X[l,1] X[l,10] }
      \hline
      F1 & By 2020, information will be used to reinvent, digitalize or eliminate 80\% of business processes and products from a decade earlier.\\ \hline
      F2 & By 2017, more than 30\% of enterprise access to broadly based Big Data will be via intermediary data broker services, serving context to business decisions.\\ \hline
      F3 & By 2017, more than 20\% of customer-facing analytic deployments will provide product tracking information leveraging the IoT\\ \hline
      \hline
    \end{tabu}
  \end{center}
  \caption{Trends within Big Data predicted by Gartner (published by \underline{F}orbes)}
  \label{tab:garntertrends}
\end{table}

Top Big Data and Analytics Trends for
2015\footnote{\url{http://www.zdnet.com/article/2015-interesting-big-data-and-analytics-trends/}}: 

\begin{table}[H]
  \begin{center}
    \begin{tabular}{l|l}
      \hline
      Z1 & More Magic\\
      Z2 & Datafication\\
      Z3 & Multipolar Analytics\\
      Z4 & Fluid Analysis\\
      Z5 & Community\\
      Z6 & Analytic Ecosystems\\
      Z7 & Data Privacy\\
      \hline
    \end{tabular}
  \end{center}
  \caption{Big Data and Analytics trends predicted in 2015 by \underline{Z}Dnet.}
  \label{tab:zdnettrends}
\end{table}

CIO's 5 Big Data Technology Predictions for
2015\footnote{\url{http://www.cio.com/article/2862014/big-data/5-big-data-technology-predictions-for-2015.html}}:

\begin{table}[H]
  \begin{center}
    \begin{tabu} to \columnwidth { X[l,1] | X[c,5]}
      \hline
      C1 & Data Agility Emerges as a Top Focus\\ \hline 
      C2 & Organizations Move from Data Lakes to Processing Data Platforms\\ \hline
      C3 & Self-Service Big Data Goes Mainstream\\ \hline
      C4 & Hadoop Vendor Consolidation: New Business Models Evolve\\ \hline
      C5 & Enterprise Architects Separate the Big Hype from Big Data\\ 
      \hline
    \end{tabu}
  \end{center}
  \caption{\underline{C}IO's Big Data Technology predictions for 2015.}
  \label{tab:ciopredictions}
\end{table}

We have enumerated the headings from the trend reports for easier
referencing.

We are aware that this is a limited subset of all available trend
reports, though these should give at least a general impression and a
basis for comparing our trend indication based on the literature
study. We have omitted reports that require registration.

Based on our literature study, we can indicate that Application is
more on the rise than Variety, Velocity and Volume, thus Big Data technology is
becoming more applied.

The latter is reflected in F1, F2, C3, C4, C5 and Z2.

%Analysis of data is mentioned in F3, Z3, Z4, and Z6. Variety is
%closely related to analytics.

Volume and Velocity do not seem to be reflected in these reports.

On general terms, we can state that the reports agree that Big Data is
becoming more mature and therefore more applied and that analytics is
the path to choose if you want to stay in front of the state-of-the
art. This is supported by Kambatla et al.~\cite{Kambatla2014}.

\subsection{Related mappings, surveys and reviews}
\label{sec:relatedsurveys}

In addition to the above, we also identified studies that did not meet
our inclusion criteria; though do provide a contribution in creating
an overview of a part of the Big Data field. Below we summarize the
type of contribution and their conclusions.

Sakr et al.~\cite{Sakr2013} provide a comprehensive survey
for a family of approaches and mechanisms of large-scale data
processing mechanisms that have been implemented based on the original
idea of the MapReduce framework and are currently gaining a lot of
momentum in both research and industrial communities. They also cover a
set of introduced systems that have been implemented to provide
declarative programming interfaces on top of the MapReduce
framework. In addition, they review several large-scale data processing
systems that resemble some of the ideas of the MapReduce framework for
different purposes and application scenarios

Gorodov and Gubarev~\cite{Gorodov2013} have done a review of methods for
visualizing data and provided a classification of visualization
methods in application to Big Data.

Ruixan~\cite{Ruixian2013} presents Bibliometrical Analysis on the Big
Data Research in China and summarizes research characteristics in
order to study Big Data in-depth development and the future
development of Big Data. They also provide reference information for
studies related to Library and Information Studies. They conclude that
research based on Big Data has taken shape though most of these papers
in the theoretical stage of exploration, lack adequate practical
support and therefore recommend to intensify efforts based on theory
and practice.

Chen and Zhang~\cite{PhilipChen2014} have done a comprehensive survey
of Big Data technologies, techniques, challenges and
applications. They offer a close view of Big Data applications
opportunities and challenges as well as techniques that is currently
adopted and used to solve Big Data problems.

Jeong and Ghani~\cite{Jeong2014} have done a review of semantic
technologies for Big Data and conclude that their analysis shows that
there is a need to put more effort into proposing new approaches, and
that tools must be created that support researchers and practitioners
in realizing the true power of semantic computing and solving the
crucial issues of Big Data.

Gandomi and Haider~\cite{Gandomi2014} present a consolidated
description of Big Data by integrating definitions from practitioners
and academics. The paper's primary focus is on the analytic methods
used for Big Data. A particular distinguishing feature of this paper
is its focus on analytics related to unstructured data, which
according to these authors constitute 95\% of Big Data.

Wang and Krishnan~\cite{Wang2014c} present a review with an objective
to provide an overview of the features of clinical Big Data. They
describe a few commonly employed computational algorithms, statistical
methods, and software tool kits for data manipulation and analysis, and
discuss the challenges and limitations in this realm.

Fern{\'a}ndez et al.~\cite{fernandez2014} focus on systems
for large-scale analytics based on the MapReduce scheme and Hadoop. They
identify several libraries and software projects that have been
developed for aiding practitioners to address this new programming
model. They also analyze the advantages and disadvantages of MapReduce,
in contrast to the classical solutions in this field. Finally, they
present a number of programming frameworks that have been proposed as
an alternative to MapReduce, developed under the premise of solving
the shortcomings of this model in certain scenarios and platforms.

Polato et al.~\cite{Polato2014} have conducted a systematic
literature review to assess research contributions to Apache
Hadoop. The objective was to identify gaps, providing motivation for
new research, and outline collaborations to Apache Hadoop and its
ecosystem, classifying and quantifying the main topics addressed in
the literature.  

Wu and Yamaguchi~\cite{Wu2014} presents a survey of Big Data
in life sciences, Big Data related projects and Semantic Web
technologies. The paper helps to understand the role of Semantic Web
technologies in the Big Data era and how they provide a promising
solution for the Big Data in life sciences.

Kambatla et al.~\cite{Kambatla2014} provide an overview of the
state-of-the-art and focus on emerging trends to highlight the
hardware, software, and application landscape of big-data analytics.

Hashem et al.~\cite{Hashem201598} have assessed the rise of big data in cloud
computing. The definition, characteristics, and classification of big
data along with some discussions on cloud computing are
introduced. The relationship between big data and cloud computing, big
data storage systems, and Hadoop technology are also
discussed. Furthermore, research challenges are investigated, with
focus on scalability, availability, data integrity, data
transformation, data quality, data heterogeneity, privacy, legal and
regulatory issues, and governance. Lastly, they give a summary of open
research issues that require substantial research efforts.

As for ongoing projects, The Byte project (EU FP7) is also investigating
the research field of Big Data. And the Big Data Value Association is
an initiative with the goal to provide the Big Data Value strategic
research agenda (SRIA) and its regular updates, defining and
monitoring the metrics of the cPPP and joining the European Commission
in the cPPP partnership board.

None of the studies above have thoroughly mapped the existing
knowledge against the Big Data V concepts, nor assessed whether the
contributions have created empirical results

%%% Local Variables:
%%% mode: latex
%%% TeX-master: "../bigdatasms"
%%% End:

\section{Discussion}
\label{sec:discussion}

There are some limitations to this study. The first limitation is that
we used a single source for our
search. Scopus\footnote{\url{www.scopus.com}} claims to be ``the
largest abstract and citation database of peer-reviewed literature:
scientific journals, books and conference proceedings'', making it a
valid choice. Scopus also returned a super set of the search results we
got from trying the same query on IEEE Xplore, ACM digital library and
Compendex. With regards to the mapping processes, one could claim that
both researchers should have read all abstracts and discussed
all. Instead, we did a pre-mapping in tandem and after that split the
work. Each was working for himself, excluding the clear outliers based
on the exclusion criteria and including the clear paper to include. In
case of even the slightest doubt, we marked the publication and
discussed these publications later on.  One can also argue that it is
a limitation to limit the mapping to the 3V's as well as application
areas and not include the other ``V's'' that appear in the papers. We
argue that sticking to the original 3 V's gives a much more concise
overview than also including non-standard V's that emerge. The 3 basic
V's are well defined, whereas others V's are open for interpretation.
Another limitation is the definition of the empirical work. However we
do not use the definitions, we just record the words used in abstracts
that are also word that describe empirical methods. If an authors
claims that they have done an experiment and that the results has been
evaluated, we noted this and did not read the full publication in order
to investigate if this is really true. We have not assessed the
quality of the work carried out in detail, other than noting that the
publication is a peer reviewed journal. A possible point of critique
of this mapping study can be that we only searched for publications
that was part of a publication that was reported as a journal as the
field of data science, analytics, databases also has a large amount of
high quality contributions disseminated through top level conferences.
However the researchers had to make a practical choice between not
being able to perform this mapping study or reducing the number of
studies to be included. Removing all studies not being part of a
journal seemed like a fair decision given our hard choice, as it does
not discrimiate across the different sub-field that contributes to Big
Data research. It can be argued that we did not find very
clear trends in the analyzed data. Finally, the method for comparing the
correlation between our result and the non-scientific trend reports
can be argued as being weak as the method of selection of these
reports were not as rigourus as the methods applied in selecting the
studies. However, the trends do coincide.

\section{Conclusion and Recommendations}
\label{sec:conclusion}
Typically a mapping study does not assess quality, though as Big Data
is and has been a very "hot topic" over the past years, the term
appears in very many papers including papers that have not contributed
to Big Data research. Therefore, we chose to only include papers that
have some form of empirical approach in order to eliminate the chance
of analyzing papers that are not contributing towards forwarding the
evidence base of Big Data research.  A total of 210 articles were
included and 151 of these have been coded against one or more of the
three ``main V's''. In addition, we have an overview of application
areas (meaning Big Data technology has been applied, though not
contributed to forwarding one of the V's).

\firstresearchquestion \textbf{Answer:} At the time of search we found
~\cite{chen2013big} and \cite{jee2013potentiality} that are labeled as
reviews, however they were not systematic. In
~\cite{park2013decomposing} Park et al. use a systematic approach; the
paper presents findings on the social networks of authors in
co-authored papers within the Big Data field.

\secondresearchquestion \textbf{Answer:} We found that on average a bit less than 10\%
(for details, refer to Table~\ref{tab:totalpublications}) of the
retrieved publications include a form of empirical approach. We also
identified the type of empirical method was used. For details see
Table~\ref{tab:empiricalmethodsmatrix} and
Table~\ref{tab:empiricalmethodsyear}. In the paper ``The Future of
Empirical Methods in Software Engineering Research'', Sj{\o}berg
et al.~\cite{sjoberg2007future} state that "an average of the reviews
indicates that about 20\% of all papers report empirical
studies". This means that the use of empirical methods in Big Data
research is below average.

\thirdresearchquestion \textbf{Answer:} We identified papers that could clearly
be classified as contributing to the Big Data field within either an application
area, or technology within Volume, Velocity or Variety. The analysis that
followed revealed that Volume (105 papers) has received the most attention from
researchers, whereas Variety (50 papers) and Velocity (22 papers) is
respectively half and a quarter of volume in number of publications. When one
looks into the deviation per year, we can see that all V's are still increasing
in absolute numbers. For more information see section \ref{sec:results} and
figure \ref{fig:venn_diagram}.

\fourthresearchquestion \textbf{Answer:} Big data is within many
different application areas, we identified 65 papers describing the
use of Big Data technology within an application area. We can see an
increase in papers addressing an application area over time. For more
details see section \ref{sec:analysis}.

\fifthresearchquestion \textbf{Answer:} From the studies retrieved by
our search, limiting to contributions that are classified as journals
by Scopus, we found that \textbf{Lecture Notes in Computer Science} is
the most prominent channel featured in our selected papers, followed
by \textbf{Proceedings of the VLDB Endowment} and \textbf{Future
  Generation Computer Systems}, for the full list see table
\ref{tab:journals}.

\sixthresearchquestion \textbf{Answer:} Based on our literature study, we can
indicate that Application is more on the rise than Variety, Velocity and Volume,
thus Big Data technology is becoming more applied. Referring to the analysis in
section 4, we can -on general terms- state Big Data is becoming more mature and
therefore more applied and that analytics is the path to choose if you want to
stay in front of the state-of-the art.

\textbf{Recommendations:} The share of publications containing empirical results
is well below the average compared to computer science research as a whole. In
order to mature the research on Big Data, we recommend to both use the evidence
base of existing empirical studies in Big Data and we recommend applying
empirical methods to strengthen the confidence in the reported results. As seen
in Table \ref{tab:totalpublications} and \ref{tab:inclusionmappingvsandapplicationareas}
all of the V's seem to be stable in their share of publications within the Big
Data field. Publications of Application of Big Data technologies is rising, a
natural explanation for this is that the Big Data field and technologies has
matured enough for applications. The least addressed areas of Big Data is
Velocity and Variety.

\begin{acknowledgements}
We would like to thank Jacqueline Floch, Tore Dyb{\aa}, Babak Farshchian and
Helge Langseth for their invaluable input and work on this paper. The work
presented in this paper is funded by BigFut project (SINTEF internally funded
project 102003299) , Norwegian University of Science and Technology and the
EXPOSED SFI (NRC grant number 6021720) project.
\end{acknowledgements}

\section{Competing interests}
The authors declare that they have no competing interests.

%%% Local Variables:
%%% mode: latex
%%% TeX-master: "../bigdatasms"
%%% End:

\bibliographystyle{spmpsci}
\bibliography{ref}

\end{document}